\documentclass[useAMS,usenatbib]{mn2e}

\usepackage{graphics,graphicx}
\usepackage{epstopdf}
\usepackage{epsfig}

\newcommand {\gtrsim} {\ {\raise-.5ex\hbox{$\buildrel>\over\sim$}}\ }
\newcommand {\lessim} {\ {\raise-.5ex\hbox{$\buildrel<\over\sim$}}\ }

\title[SN~2011ay]{The early phases of the type Iax supernova SN~2011ay}

\author[Szalai et al.]{Tam\'{a}s Szalai$^{1}$, J\'{o}zsef Vink\'{o}$^{1,2}$, Kriszti\'{a}n S\'{a}rneczky$^{3,4}$, Katalin Tak\'{a}ts$^{1,5}$
\newauthor J\'{o}zsef M. Benk\H{o}$^{3}$, J\'{a}nos Kelemen$^{3}$, Zolt\'{a}n Kuli$^{3,6}$, Jeffrey M. Silverman$^{2,7}$,
\newauthor G. Howie Marion$^{2,8}$, and J. Craig Wheeler$^{2}$ \\
\\
$^{1}$Department of Optics and Quantum Electronics, University of Szeged, D\'om t\'er 9., Szeged H-6720, Hungary\\
$^{2}$Department of Astronomy, University of Texas at Austin, 1 University Station C1400, Austin, TX 78712-0259, USA\\
$^{3}$Konkoly Observatory, MTA CSFK, Konkoly Thege u. 15-17, H-1121 Budapest, Hungary\\
$^{4}$ELTE Gothard-Lend\"{u}let Research Group, 9700 Szombathely, Hungary\\
$^{5}$Departamento Ciencias F\'{i}sicas, Universidad Andres Bello, Av. Rep\'{u}blica 252, Santiago, Chile\\
$^{6}$Hungarian Astronomical Association, Budapest, Hungary\\
$^{7}$NSF Astronomy and Astrophysics Postdoctoral Fellow\\
$^{8}$Harvard-Smithsonian Center for Astrophysics, 60 Garden Street, Cambridge, MA 102138, USA}

\begin{document}

\date{in original form 2014}

\pagerange{\pageref{firstpage}--\pageref{lastpage}} \pubyear{2013}

\maketitle

\label{firstpage}

\begin{abstract}

We present a detailed study of the early phases of the peculiar supernova 2011ay based on BVRI photometry obtained at Konkoly Observatory, Hungary, and optical spectra taken with the Hobby-Eberly Telescope at McDonald Observatory, Texas. 
The spectral analysis carried out with SYN++ and SYNAPPS confirms that 
SN~2011ay belongs to the recently defined class of SNe Iax, which is also supported by the properties of its light and color curves.  
The estimated photospheric temperature around maximum light, $T_{phot} \sim$ 8,000 K, is lower than in most Type Ia SNe, 
which results in the appearance of strong Fe\,\begin{small}II\end{small} features in the spectra of SN~2011ay, even 
during the early phases. 
We also show that strong blending with metal features (those of Ti\,\begin{small}II\end{small}, 
Fe\,\begin{small}II\end{small}, Co\,\begin{small}II\end{small}) makes the direct analysis of the broad spectral features very difficult, and this may be true for all SNe Iax. We find two alternative spectrum models that both describe the
observed spectra adequately, but their photospheric velocities differ by at least $\sim 3,000$ km s$^{-1}$.
The quasi-bolometric light curve of SN~2011ay has been assembled by integrating the UV-optical spectral 
energy distributions. Fitting a modified Arnett-model to $L_{bol}(t)$, 
the moment of explosion and other physical parameters, i.e. the rise time to maximum, the $^{56}$Ni mass and the total ejecta mass are estimated as $t_{rise} \sim$14$\pm$1 days, $M_{Ni} \sim$0.22$\pm$0.01 $M_{\odot}$ and $M_{ej}\sim$ 0.8 $M_{\odot}$,
respectively.

\end{abstract}


\section{Introduction}\label{intro}

Type Iax supernovae (SNe~Iax), previously labeled as SN~2002cx-like ones, are 
spectroscopically similar to SNe Ia, but thought to have lower velocities at maximum light and lower peak magnitudes \citep{Foley13}. 
There are 25 members of this class known to date, but it is suggested that in a given volume there may be 30-40 SNe Iax for every 100 SNe Ia.  

The major question about SNe Iax is the nature of their progenitors. 
The similarity of their light curves to those of ``normal'' SNe Ia (especially their rise times), their spectral features, and  
the lack of X-ray and radio detections suggest that they originate from compact progenitors, probably white dwarfs (WDs). 
At the same time, explosions causing the total disruption of WDs 
(either via detonation or via pure deflagration) are not in agreement with the observational results.
This statement was recently strengthened by \citet{McCully14a} who analyzed late-time spectra 
of two SNe Iax.

Partial disruption of WDs seems to be {a feasible} model to explain the properties of 
this new class of stellar explosions \citep{Foley13,McCully14a}. This model seems to give a plausible explanation 
for the diversity of this class (including extremely faint and low-velocity explosions, like SN~2008ha). 
It is also supported by recently published results of numerical 3D simulations \citep{Jordan12,Kromer13}.
As \citet{Foley13} and \citet{Wang13} proposed, WDs accreting material from their helium-star companion in binary systems may be good candidates 
for the progenitors of SNe Iax. This assumption has been recently strengthened by 
the detection of the luminuous, blue progenitor system of the type Iax SN 2012Z \citep{McCully14b}.
At the same time, the observed properties of this object, and maybe those of other SNe Iax, can be also explained with the pulsational delayed 
detonation of a single WD \citep[see][and references therein]{Stritzinger15}.

Other ideas, e.g. the theory of the ``.Ia'' events \citep[which are thought to be luminous helium shell flashes of WDs 
in binary systems with low accretion rates,][]{Bildsten07}, do not seem to give self-consistent
description of the observations \citep{Foley09,Foley13}. Note, however, that the .Ia models might be good 
alternatives to explain the extremely low-luminosity events, like SN~2008ha.

On the other hand, so far almost all SNe Iax have been found in late-type galaxies with recent, strong star formation 
activity \citep{Lyman13,Foley13}. This fact might lead to the suspicion that SNe Iax actually could be low-luminosity core-collapse 
(CC) events emerging from 7-9 M$_{\odot}$, stripped-envelope stars \citep{Valenti09,Lyman13}, even though 
the observed properties of SNe Iax are more similar to those of Ia than CC SNe. 
In this scenario the explosion might not have enough energy to eject the entire  
stellar envelope, thus, part of the ejecta falls back to the central remnant \citep{Valenti09,Moriya10}.
Note, however, that SN~2008ge, another SN Iax, is situated in a region that has no detectable star-formation, 
thus, there are probably no massive stars near the SN site \citep{Foley10b,Foley13}.

To understand the general properties of SNe Iax, it is important to study single, well-observed objects as 
thoroughly as possible. Up to now, detailed photometric and spectroscopic analyses have been published only on a few of them: SN~2002cx \citep{Li03,Branch04,Jha06}, 
SN~2005hk \citep{Chornock06,Phillips07,Sahu08,McCully14a}, 
SN~2007qd \citep{McClelland10}, 
SN~2008A \citep{McCully14a}, 
SN~2008ge \citep{Foley10b}, 
SN~2008ha \citep{Foley09,Valenti09,Foley10a}, 
SN~2009ku \citep{Narayan11}, 
SN~2010ae \citep{Stritzinger14},
SN~2012Z \citep{McCully14b,Stritzinger15,Yamanaka15},
and SN~2014dt \citep{Foley15}.
Moreover, \citet{Foley13} presented light curves and spectra of some other SNe Iax for which detailed studies are in progress.

Here we present a detailed analysis of SN~2011ay, another SN Iax.
This object was discovered on March 18.18 UT by the KAIT/LOSS program with an apparent, unfiltered brightness of 17.7 mag \citep{Blanchard11}. 
The SN located 9.3\arcsec east and 1.4\arcsec south from the center of NGC~2315. The average Hubble-flow distance of the host galaxy is 
86.9$\pm$6.9 Mpc, as listed in the NASA/IPAC Extragalactic Database (NED). Based on a quick analysis of an early spectrum, \citet{Pogge11} 
classified SN~2011ay as a peculiar Ia, while \citet{Silverman11} reported that it may belong to the SN 2002cx-like events. \citet{Foley13} 
confirmed the classification and also published some optical photometry and 
spectra of this object.

Our motivation was to carry out a detailed analysis using data obtained during the early photospheric phase of the SN. 
We present our simultaneous photometric and spectroscopic observations in Section \ref{obs}, while the light and color 
curves, the UV-optical spectral energy distributions (SEDs) and the results of modeling the spectra and the light curve 
are shown in Section \ref{anal}. In Section \ref{conc} we discuss our results and present our conclusions.

\section{Observations and data reduction}\label{obs}

\subsection{Photometry}\label{obs_phot}

\begin{figure}
\begin{center}
\leavevmode
\includegraphics[width=8cm]{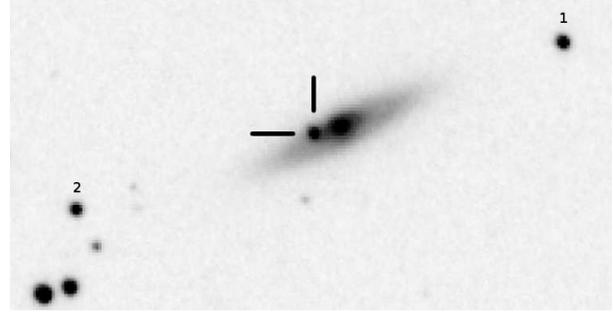}
\end{center}
\caption{BVI composite image of the SN~2011ay and its host galaxy on 2011 March 24 (0.6/0.9 m Schmidt-telescope, Konkoly Observatory, Hungary). The numbered objects 
were used as local comparison stars.}
\label{fig:sn_img}
\end{figure}

Ground-based photometric observations for SN~2011ay were obtained from the Piszk\'{e}stet\H{o} Mountain Station of Konkoly Observatory, Hungary. We used the 0.6/0.9 m Schmidt-telescope and the 1.0m RCC-telescope, both equipped with Bessell 
BVRI filters. Figure \ref{fig:sn_img} shows a BVI composite image of the SN and its host galaxy.

We used the {\it daophot/allstar} task in IRAF\footnote{IRAF is distributed by the National Optical Astronomy Observatories, which are operated by the Association of Universities for 
Research in Astronomy, Inc., under cooperative agreement with the National Science Foundation.} to carry out PSF-photometry on the SN and two nearby comparison stars (see Figure \ref{fig:sn_img}). Because the SN appeared close to the center of the nearly edge-on host galaxy, the value of background flux level was estimated interactively in every case. The background level
estimates produced by the {\it fitskypars} task were adjusted by eye and the background was subtracted manually for each frame. 

The instrumental magnitudes were transformed to the standard system applying the following equations:

\begin{eqnarray}
V - v = C_V \cdot (V-I) + \zeta_V \nonumber\\
(B-V) = C_{BV} \cdot (b-v) + \zeta_{BV} \nonumber\\
(V-R) = C_{VR} \cdot (v-r) + \zeta_{VR} \nonumber\\
(V-I) = C_{VI} \cdot (v-i) + \zeta_{VI},
\label{eq:standtr}
\end{eqnarray}

\noindent where lowercase and uppercase letters denote instrumental and standard magnitudes, respectively.
The color terms ($C_X$), determined by measuring Landolt standard stars during photometric conditions close to the epochs of the SN observations, were the followings: 
C$_V$ = $−$0.047, 
C$_{BV}$ = 1.228, 
C$_{VR}$ = 0.960, 
C$_{VI}$ = 0.934. 
Because simultaneous observation of the Landolt field and the SN field was not possible, we applied 
Eq. (\ref{eq:standtr}) to the differential magnitudes $\Delta m = m_{SN} - m_{C1,2}$ constructed from the
instrumental magnitudes of the SN and the comparison stars, thus, eliminating the zero points ($\zeta_X$).
Then, using the standard BVRI magnitudes of the comparison stars calculated from their SDSS-magnitudes \citep[based on the calibration of][]{Jordi06}, 
the standard magnitudes of SN~2011ay were obtained in the Johnson-Cousins system.
The results are presented in Table \ref{tab:mag}. Errors (given in parentheses) contain both the uncertainties of 
the PSF-photometry and the standard transformation.

The optical data were supplemented by the available {\it Swift/UVOT} data (reduced using standard HEAsoft tasks). Individual frames were summed with the {\it uvotimsum} task, and
magnitudes were determined via aperture photometry using the task {\it uvotsource}. 
The object could be roughly separated from the galactic nucleus only in channels b, u, and uw1, while it got too faint after $\sim$2 weeks. The results of {\it Swift}-photometry are 
shown in Table \ref{tab:smag}.

\begin{table}
\begin{center}
\caption{BVRI magnitudes of SN~2011ay from the Konkoly Observatory, Hungary}
\label{tab:mag}
\begin{tabular}{@{}c|cccc@{}}
\hline
\hline
JD$^a$ & B (mag) & V (mag) & R (mag) & I (mag) \\
\hline
5645.3 & 16.89(.07) & 16.97(.05) & 16.75(.06) & 16.74(.10) \\
5647.3 & 16.85(.02) & 16.71(.05) & 16.52(.06) & 16.33(.08) \\
5649.3 & 16.99(.03) & 16.62(.07) & 16.44(.06) & 16.31(.08) \\
5650.3 & 16.99(.06) & 16.54(.04) & 16.50(.05) & 16.28(.08) \\
5651.3 & 16.97(.05) & 16.65(.07) & 16.42(.06) & 16.21(.07) \\
5655.3 & 17.30(.06) & 16.62(.07) & 16.31(.06) & 16.10(.07) \\
5657.3 & 17.38(.07) & 16.63(.06) & 16.28(.06) & 16.09(.07) \\
5661.6 & 18.02(.15) & 16.86(.06) & 16.36(.05) & 16.21(.07) \\
5669.3 & 18.82(.23) & 17.71(.03) & 16.88(.04) & 16.43(.02) \\
5671.3 & 19.05(.12) & 17.78(.09) & 17.05(.03) & 16.48(.02) \\
\hline
\end{tabular}
\end{center}
\smallskip
{\bf Notes.} $^{(a)}$JD$-$2,450,000. Errors are given in parentheses.
\end{table}

\begin{table}
\begin{center}
\caption{Swift magnitudes of SN~2011ay}
\label{tab:smag}
\begin{tabular}{@{}c|ccc@{}}
\hline
\hline
JD$^a$ & UVW1 (mag) & U (mag) & B (mag) \\
\hline
5645.5 & 17.70(.12) & 16.53(.07) & 16.59(.05) \\
5648.0 & 17.98(.11) & 16.63(.07) & 16.66(.05) \\
5650.1 & 18.22(.13) & 16.75(.06) & 16.67(.05) \\
5651.8 & 18.32(.14) & 16.91(.08) & 16.80(.05) \\
5653.5 & 18.17(.13) & 17.14(.10) & 16.79(.06) \\
5655.6 & 18.66(.25) & 17.26(.12) & 17.01(.08) \\
5657.6 & 18.73(.19) & 17.36(.10) & 17.18(.07) \\
5660.1 & 18.83(.22) & 17.88(.18) & 17.54(.11) \\
\hline
\end{tabular}
\end{center}
\smallskip
{\bf Notes.} $^{(a)}$JD$-$2,450,000. Errors are given in parentheses.
\end{table}

\subsection{Spectroscopy}\label{obs_sp}

\begin{figure}
\begin{center}
\leavevmode
\includegraphics[width=8cm]{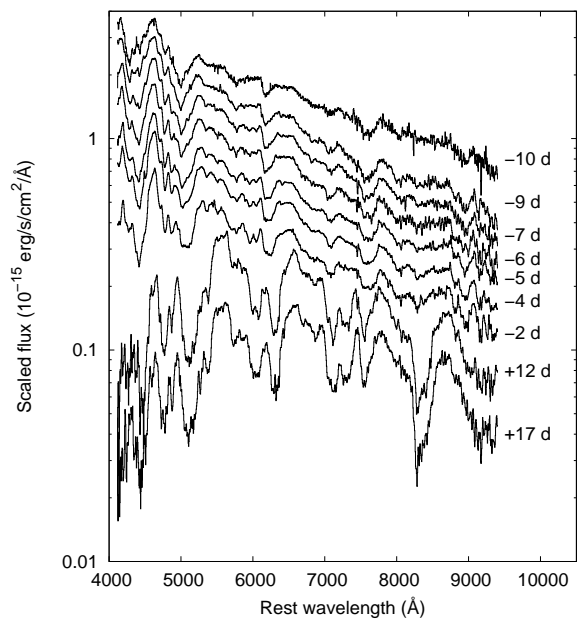}
\end{center}
\caption{HET spectra of SN~2011ay. Epochs are given relative to V-band maximum. The spectra are shifted vertically for better visibility.}
\label{fig:spectra}
\end{figure}

Optical spectra of SN~2011ay were obtained with the 9.2m Hobby-Eberly Telescope (HET) Marcario Low Resolution Spectrograph \citep[LRS,][]{Hill98} at McDonald Observatory, Texas, between March 24 and April 19, 2011. LRS has a spectral coverage 
of 4,200--10,200 \AA\ and a resolving power of $\lambda / \Delta\lambda \sim$ 600. The data were reduced with standard IRAF routines. Table \ref{tab:sp} contains the journal of the spectroscopic observations, while the extracted, wavelength- and flux-calibrated spectra are collected in Figure \ref{fig:spectra}.

\begin{table}
\begin{center}
\caption{Log of spectral observations obtained with HET LRS}
\label{tab:sp}
\begin{tabular}{cccc}
\hline
\hline
Date &  JD$^a$ & Phase$^b$ (d) & Exp. time (s)\\
\hline
2011-03-23 & 5643.2 & $-$10 & 1500 \\
2011-03-24 & 5644.2 & $-$9 & 1800 \\
2011-03-26 & 5646.2 & $-$7 & 1500 \\
2011-03-27 & 5647.2 & $-$6 & 1500 \\
2011-03-28 & 5648.2 & $-$5 & 1500 \\
2011-03-29 & 5649.2 & $-$4 & 1500 \\
2011-03-31 & 5651.2 & $-$2 & 1500 \\
2011-04-14 & 5665.1 & +12 & 1500 \\
2011-04-19 & 5670.1 & +17 & 1500 \\
\hline
\end{tabular}
\end{center}
\smallskip
{\bf Notes.} $^{(a)}$JD$-$2,450,000; $^{(b)}$ with respect to the moment of maximum light in V-band.
\end{table}

\section{Data analysis and results}\label{anal}

\subsection{Light curves and colors}\label{ana_lc}

\begin{figure}
\begin{center}
\leavevmode
\includegraphics[width=8cm]{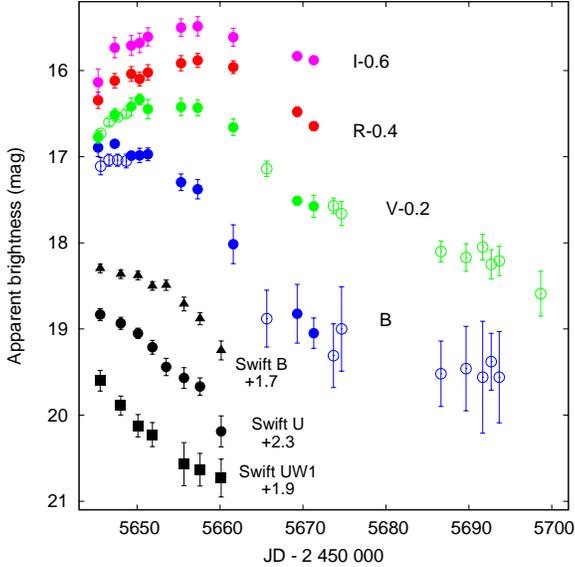}
\end{center}
\caption{Standard BVRI (Konkoly Observatory) and Swift light curves of SN 2011ay (not corrected for reddening). The open circles denote the B and V data from \citet{Foley13}.}
\label{fig:lc}
\end{figure}

The light curves consisting of standard BVRI and Swift magnitudes are shown in Figure \ref{fig:lc}. The Konkoly B and V curves are generally consistent with the ones published 
by \citet{Foley13}, but the sampling of our data around maximum is better (at longer wavelengths they observed SN~20011ay with $r$ and $i$ filters). 
As seen in Fig.~\ref{fig:lc}, the data by \citet{Foley13} do not provide new information in addition to our photometry during the early phases. While they followed the SN longer, their late-time photometry is much more uncertain, probably due to the increasing contribution from the host galaxy
background flux. Thus, for further analysis we used only our data, but note that the results are consistent
with the photometry by \citet{Foley13}.

By fitting low-order polynomials to the data around the maxima, the light-curve parameters could be
determined with relatively low uncertainties. These parameters are listed in Table \ref{tab:photpar}.
Note that for normal SNe Ia the moment of B maximum is usually used as a reference epoch. 
Instead, we used the time of V maximum, which is closer to the moment of maximum luminosity, 
to be consistent with the analysis presented by \citet{Foley13}. 

\begin{table*}
\begin{center}
\caption{Light-curve parameters of SN~2011ay}
\label{tab:photpar}
\begin{tabular}{@{}l|cccc@{}}
\hline
\hline
~ & B & V & R & I \\
\hline
$t_{max}$ & ~ & ~ & ~  \\
(JD-2,450,000) & 5646.6(1.1) & 5653.6(0.4) & 5657.2(1.1) & 5656.2(0.8)\\
Peak mag. & 16.89(.08) & 16.56(.08) & 16.30( .07) & 16.08(.08) \\
Peak abs. mag. & $-$18.15(.17) & $-$18.39(.18) & $-$18.60(.17) & $-$18.76(.18) \\
$\Delta m_{15}$ (mag) & 1.11(.16) & 0.95(.08) & 0.82(.07) & 0.40(.08) \\
\hline
\end{tabular}
\end{center}
\smallskip
{\bf Notes.} Errors are given in parentheses.
\end{table*}

\begin{figure}
\begin{center}
\leavevmode
\includegraphics[width=8cm]{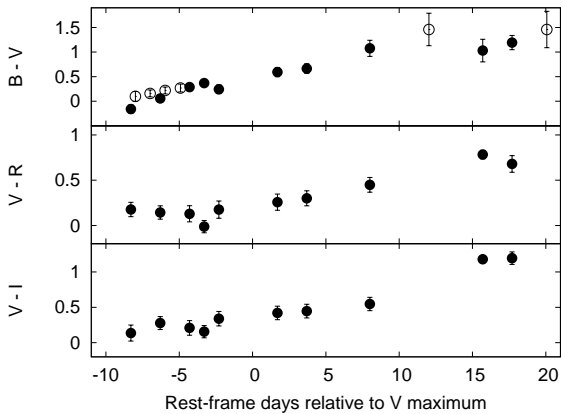}
\end{center}
\caption{Color curves of SN~2011ay (corrected for Milky Way reddening, see the text for details). 
The values plotted with open circles were calculated using the B and V datapoints of \citet{Foley13}.}
\label{fig:color}
\end{figure}

The color evolution of SN~2011ay (Figure \ref{fig:color}) is very similar to other SNe Iax \citep[see e.g.][]{Li03,Phillips07,Foley13}: it gets redder after V-maximum until it reaches a nearly constant color at 
15-20 days after maximum.
Although the interstellar matter (ISM) in the host galaxy may contribute to the total reddening, 
it is probably low, because the SN appeared very blue, $(B-V) < 0$ mag, at the earliest observed epochs. 
Thus, the color curves have been corrected only for Milky Way reddening, 
using $E(B-V)$ = 0.081 mag, $A_V$ = 0.23 mag, $A_R$ = 0.18 mag, and $A_I$ = 0.13 mag \citep{Schlafly11}.

\subsection{Spectroscopic analysis}\label{ana_sp}

In the following, we present detailed analysis of some of the HET spectra shown in Fig.~\ref{fig:spectra} and
Table~\ref{tab:sp}. The $-9$d, $-2$d and $+17$d spectra\footnote{We use the moment of V-band maximum light for assigning
phases to the observed spectra, in accord with \citet{Foley13}} were selected for this purpose, because they have the 
highest signal-to-noise and show the largest differences in their spectral features, i.e. the most noticeable evolution.
Moreover, two additional spectra obtained by \citet{Foley13} were also included in the sample of analyzed spectra.
These two spectra were taken with the Lick/Kast spectrograph on April 2 and 5, 2011,
at $0$d and $+3$d phases, respectively.

The application of the Supernova Identification (SNID) code \citep{Blondin07} to the pre-maximum spectra
of SN~2011ay revealed some similarity with spectra of normal Ic SNe. However, using SNID for the $+17$d HET
spectrum we found that the relatively weak Si, dominating Fe\,\begin{small}II\end{small} and 
strong Ca\,\begin{small}II\end{small} features clearly imply that SN~2011ay 
most closely resembles SN 2005hk \citep[see e.g.][]{Sahu08}, i.e. it belongs to the Iax 
subclass, as first pointed out by \citet{Silverman11} and \citet{Foley13}. 

For more quantitative analysis, we applied the SYN++ and SYNAPPS codes \citep{Thomas11}, which are
based on the original SYNOW code \citep{JB90, Hatano99}, to model the 
selected spectra in order to reveal the chemical composition and other physical properties of the 
ejecta of SN~2011ay. 
SYNOW/SYN++/SYNAPPS assumes a fully opaque photosphere located at a specific velocity 
$v_{phot}$, where the electron-scattering optical 
depth $\tau_{es} = 1$ \citep{Hatano99}, in homologously expanding ejecta. 
The spectral features are assumed to be due to pure resonant scattering of the photons originating from the
(assumed) blackbody emission of the photosphere. The scattering of photons take place 
in the partly transparent ejecta above the photosphere, where the strong velocity gradient
(due to the rapid, homologous expansion) makes the scattering regions appear as thin planes 
perpendicular to the line of sight for each photon frequency/wavelength. This 
Sobolev-approximation, among others, is clearly a limitation for the applicability of
such a code. Nevertheless, it was found to be surprisingly effective and useful for 
modeling P~Cygni line profiles and identifying SN features \citep[see e.g.][]{Parrent14-review, Parrent14}.  

Before the fitting, all spectra were corrected 
to Milky-Way extinction (adopting $E(B-V)=0.081$ mag as above) and the redshift of the host galaxy 
using $z$ = 0.021 \citep[][]{Miller01}.


Since parametrized models, like those from SYNAPPS/SYN++, depend heavily on the {\it a-priori} identification
of ions and key parameters such as the velocity at the photosphere, we built two alternative models based
on slightly different assumptions. 

In the first model (hereafter Model A) all features were assumed as photospheric, i.e. the
line formation region for all ions extends from the top of the ejecta, parametrized as 
a fixed $v_{max}$ = $40,000$ km s$^{-1}$ at all phases, down to the sharp edge of the 
photosphere, given as $v_{phot}$ in velocity coordinates. The density structure of the
ions distributed within the line forming region is treated as a simple exponential, defined directly for the optical depth of a given feature as 
\begin{equation}
\tau(v) ~=~ \tau(v_{ref}) \times \exp[-(v - v_{ref}) / v_e],
\end{equation}
where $v_{ref}$ is a fixed reference velocity (arbitrary, but set close to $v_{phot}$) and
$v_e$ is the e-folding width of the optical depth profile. SYN++ uses a quasi-LTE approximation
for computing the relative strengths of the features for the same ion: the model specifies a single
$\tau(v_{ref})$ optical depth for a pre-selected ``reference'' feature of the given atom/ion
\citep[see e.g.][for the list of reference lines for every ion]{Hatano99}, then the optical 
depths of all other same-ion features are computed assuming Boltzmann-excitation. The excitation
temperature $T_{exc}$ can be set differently for every ion, mimicking a non-LTE-like excitation,
although it is clearly very far from a true, self-consistent non-LTE treatment of the problem.
Thus, in Model A, all features are assumed to be formed down to the photosphere, each ion can
have a different excitation temperature ($T_{exc}$), a different reference line optical depth ($\tau(v_{ref})$)
and different scale height of its line forming region ($v_e$). 

The alternative model (hereafter Model B) was built 
by assuming two essential differences from Model A: instead of requiring that all features start
to form at $v_{phot}$, the minimum velocity of their line-forming region, $v_{min}$,
may be at higher velocities than $v_{phot}$. Such features, having $v_{min} > v_{phot}$, are
called ``detached'' \citep{JB90}. This model may be more flexible than Model A, because $v_{min}$
can be different for every ion, but also less constrained, because the increasing number of 
parameters may decrease the uniqueness of the best-fit model. In addition, since $v_{phot}$ may become
much less constrained in this model, we set the initial value of $v_{phot}$ at $\sim 6,000$ 
km s$^{-1}$ as estimated by \citet{Silverman11} and \citet{Foley13} for SN~2011ay particularly 
from the absorption minimum of the feature around $\sim 6200$ \AA, which was thought to be due to 
Si\,\begin{small}II\end{small} (but see Sect.\ref{ana_vel} for a discussion of this issue).

The chemical composition of both models was assembled based on the major
features found in the spectra of other SNe Iax, namely 2002cx \citep{Branch04}, 2005hk 
\citep{Chornock06, Sahu08, McCully14a}, 2008A \citep{McCully14a}, 2008ha and 2010ae \citep{Stritzinger14}:
O\,\begin{small}I\end{small}, O\,\begin{small}II\end{small}, Na\,\begin{small}I\end{small}, 
Mg\,\begin{small}II\end{small}, Si\,\begin{small}II\end{small}, 
S\,\begin{small}II\end{small}, Ca\,\begin{small}II\end{small}, Ti\,\begin{small}II\end{small},
Fe\,\begin{small}II\end{small}, Fe\,\begin{small}III\end{small} and Co\,\begin{small}II\end{small}.
Not all ions were detected in all spectra, e.g. the ones with higher ionization potential 
(O\,\begin{small}II\end{small} and Fe\,\begin{small}III\end{small}) were not found in the 
late-phase $+17$d spectrum. 

For both models, the optimum set of parameters that give the best fit for the synthesized spectrum
to the observed one were found automatically by SYNAPPS via $\chi^2$-minimization. The 
total number of the optimized parameters were $\sim 50$ in a typical run. Note that there is an 
essential degeneracy between $\tau(v_{ref})$, $v_{phot}$ (or $v_{min}$) and $v_e$ \citep{JB90, Parrent14}, thus, their finally adopted values may not represent 
a unique solution for any spectra. 

\begin{figure*}
\begin{center}
\leavevmode
\includegraphics[width=14cm]{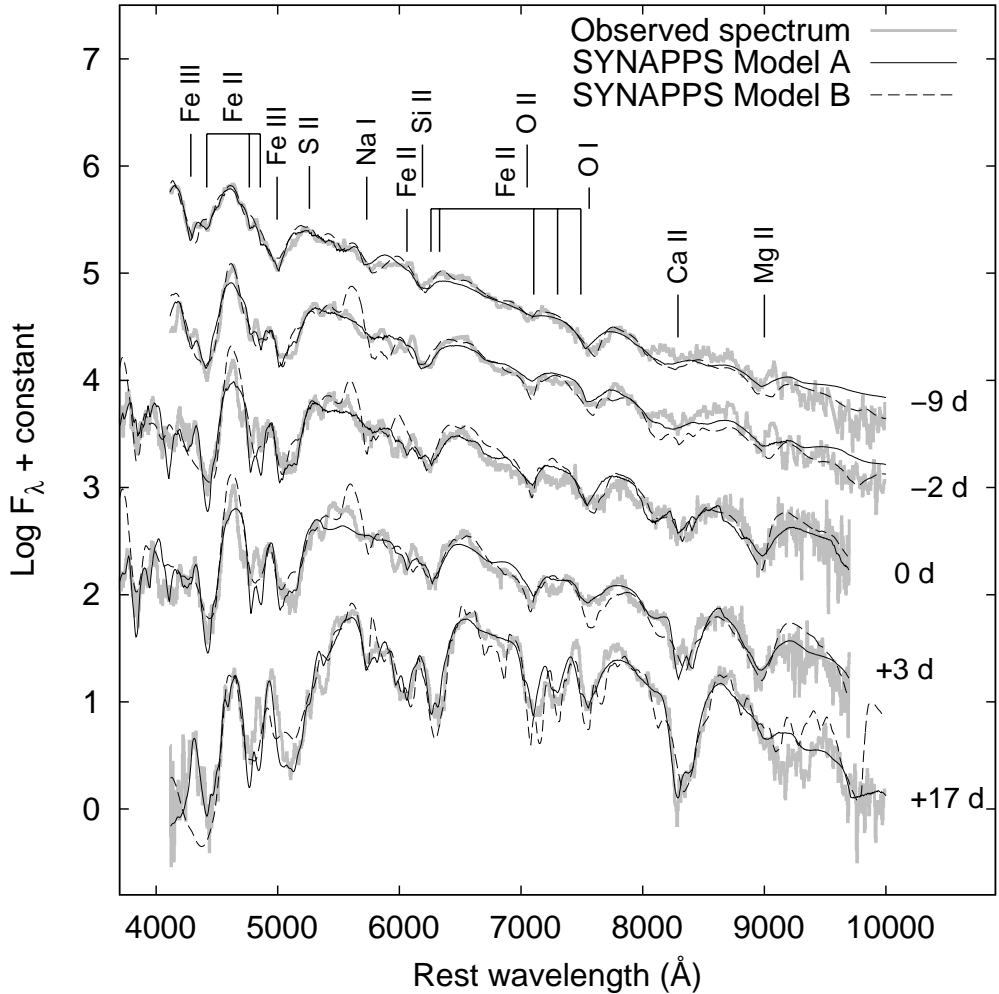}
\end{center}
\caption{The best-fit SYNAPPS model spectra for Model A (solid line) and Model B (dashed line) plotted together with the observed spectra at $-9$d, $-2$d, $0$d, $+3$d and $+17$d phases. 
Strong blending makes feature identification ambiguous at all phases. Some of the strongest contributing ions are marked by vertical lines and labels.}
\label{fig:sp-modelall}
\end{figure*}

\begin{figure*}
\begin{center}
\leavevmode
\includegraphics[width=14cm]{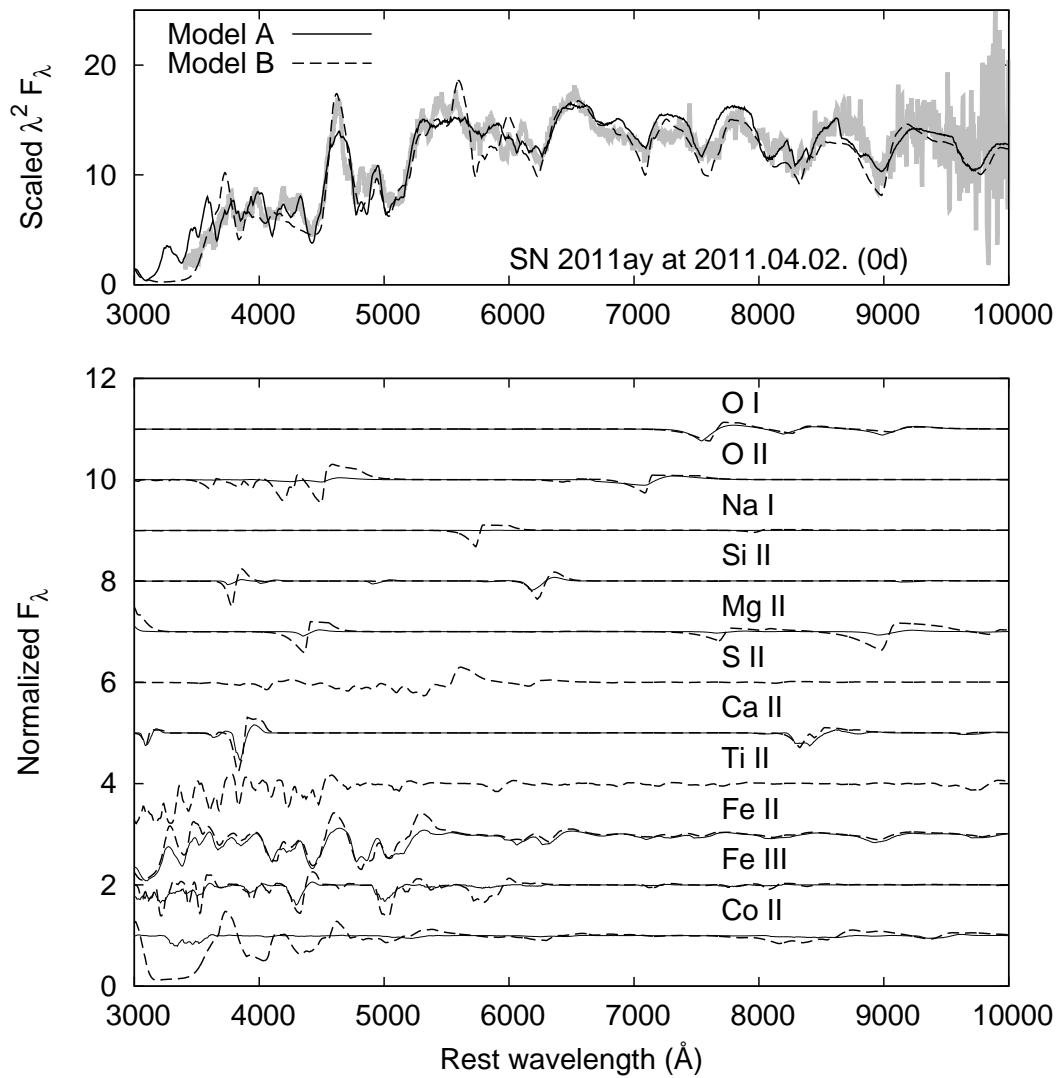}
\end{center}
\caption{Single-ion contributions to the 0d spectrum fit (solid line = Model A, dashed line = Model B).}
\label{fig:sp-modelions}
\end{figure*}

\begin{table}
\begin{center}
\caption{The photospheric velocities and temperatures for Model A and B, as found by SYNAPPS}
\label{tab:synpar}
\begin{tabular}{cccccc}
\hline
\hline
Date & Epoch & $v_{phot}^A$ & $T_{phot}^A$ & $v_{phot}^B$ & $T_{phot}^B$ \\
~ & (days) & (km s$^{-1}$) & (K) & (km s$^{-1}$) & (K) \\
\hline
March 24 & $-9$ & 9830 & 8710 & 6500 & 10,000 \\
March 31 & $-2$ & 9640 & 7460 & 6500 & 10,000 \\
April 02 & 0 & 9280 & 7260 & 6000 & 8760 \\
April 05 & +3 & 9090 & 6000 & 6500 & 8620 \\
April 19 & +17 & 8870 & 5300 & 4100 & 6490 \\
\hline
\end{tabular}
\end{center}
\smallskip
\end{table}

The final best-fit model spectra for both Model A and Model B are plotted together with the 
observations in Fig.~\ref{fig:sp-modelall}. Fig.~\ref{fig:sp-modelions} shows the single-ion
contributions to the best-fit models for the +0d spectrum, i.e. at the phase of V-band maximum. 
The basic global parameters for both models are collected in Table~\ref{tab:synpar}.

From Table~\ref{tab:synpar} and Figs~\ref{fig:sp-modelall} - \ref{fig:sp-modelions} it is 
immediately apparent that, despite having quite different $v_{phot}$ values, both Model A and B
give acceptable fits to the observed spectra. Both models reproduce the major
spectral features well, although neither fits perfectly all the weak, narrow humps 
that become stronger in the post-maximum spectra. This is somewhat surprising, because
the two models converged to $v_{phot}$ values that are different by
$\sim 3000$ km s$^{-1}$. Since the uncertainty of the photospheric velocities derived by SYNAPPS
are thought to be $\sim 500$ - $1000$ km s$^{-1}$, at least for SNe Ia \citep{Parrent14},
such a high level of ambiguity, at first, seems to be unexpected. It is, however, not
unprecedented among other SNe Iax: for example, \citet{Stritzinger14} found similar inconsistency 
between their SYNAPPS models for the simulteneous optical ($v_{phot} \sim 4200$ km s$^{-1}$) 
and near-IR ($v_{phot} \sim 1700$ km s$^{-1}$) spectra of SN~2010ae at +18d phase.  

A possible (and likely) explanation for this issue is that all modeled SN~2011ay spectra consist of heavily blended features, even at very early phases. 
There are no individual, unblended, single-ion features visible that can be identified unambiguously, 
except maybe O\,\begin{small}I\end{small} $\lambda7775$
in the pre-maximum spectra and the Ca\,\begin{small}II\end{small} near-IR triplet after maximum. In
both models considered here the presence of iron (both Fe\,\begin{small}II\end{small} and 
Fe\,\begin{small}III\end{small}) can be found at $-9$d, and blending with Fe\,\begin{small}II\end{small}
and Co\,\begin{small}II\end{small} becomes excessively dominant at later phases, after maximum light.
As noted by \citet{Branch04}, the ``iron curtain'' of strong Fe\,\begin{small}II\end{small} 
prevented the secure identification of many other features in the post-maximum spectra of SN~2002cx, and
this is exactly what we found here for SN~2011ay. The projected Doppler-velocities of
blended features become quickly ill-constrained with the increasing number of overlapping lines from
different ions.

\begin{figure}
\begin{center}
\includegraphics[width=8cm]{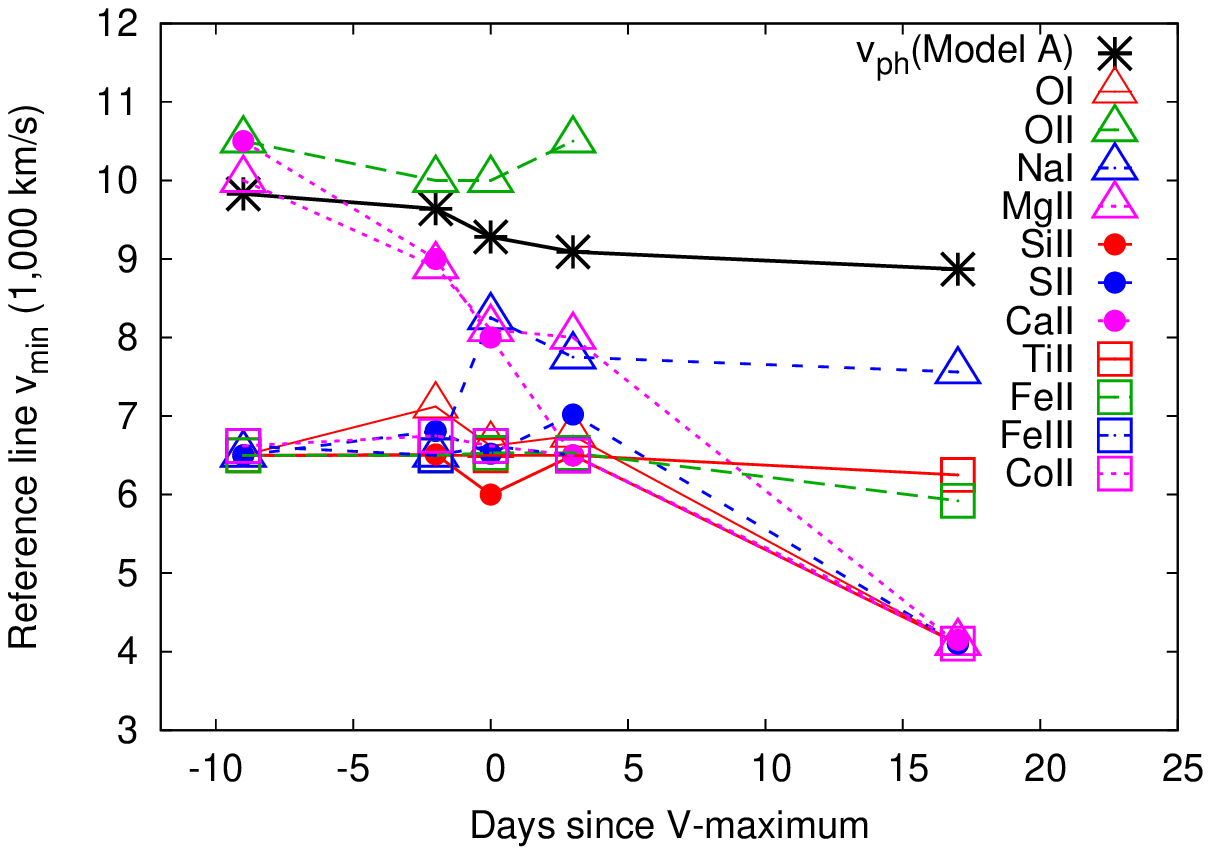}
\end{center}
\caption{Evolution of the minimum velocities of the line-forming regions for
each species in Model B. The uncertainty of these velocities is at least $\sim 1,000$ km s$^{-1}$. 
The lowest velocity at each phase corresponds to the photospheric velocity at the given epoch. For comparison, the photospheric 
velocity in Model A is also plotted with asterisks.}
\label{fig:minvel}
\end{figure}

The evolution of the minimum velocities of the line forming region for each species in Model B are plotted in Fig. \ref{fig:minvel}.
On the contrary, Model A assumed that all features are formed with a minimum velocity of $v_{phot}$ as given in Table~\ref{tab:synpar}.
It is apparent that while in Model B the photospheric velocity is significantly
lower than that in Model A, Model B contains some low-mass ions 
(O\,\begin{small}II\end{small}, Na\,\begin{small}I\end{small}, Mg\,\begin{small}II\end{small})
that are formed at higher velocities. The minimum velocities of these ``detached'' 
features are close to $v_{phot}$ of Model A, but they tend to decrease below that,
and converge toward the photosphere at later phases 
(except for O\,\begin{small}II\end{small}, which stays constantly 
at high velocity). \citet{Branch04} also found similar ``high-velocity''
features in the spectra of SN~2002cx, but only at post-maximum phases. 
Fig. \ref{fig:minvel} suggests that such detached features forming at
higher velocities may have appeared in SN~2011ay as early as 9-10 days
before maximum.

From Fig.~\ref{fig:sp-modelall} and \ref{fig:sp-modelions} it is seen that 
low-Z elements, such as O\,\begin{small}I\end{small}, Na\,\begin{small}I\end{small},
Mg\,\begin{small}II\end{small}, start appearing at very early phases 
(more than one week pre-maximum), and they remain detectable even at post-maximum. 
S\,\begin{small}II\end{small} may also be present (at least in Model B), 
which is often cited as a sign of thermonuclear explosion. 
These features are frequently observed in SNe Ia, but they are absent in SNe Ib/c.

\citet{Foley13} reported the possible presence of carbon in SN~2011ay via the detection of
C\,\begin{small}II\end{small} $\lambda6580$ and $\lambda7234$ features,
even though these features appear very weak \citep[see Fig. 23 in][]{Foley13}.
We have checked the presence/absence of C\,\begin{small}I\end{small},
C\,\begin{small}II\end{small} or C\,\begin{small}III\end{small} using SYNAPPS.
The $-9$d and $-2$d spectra were selected for this purpose, because carbon
features are expected to be the strongest during the pre-maximum phases.
For this test we adopted Model B, because that model has similar velocities 
to those adopted by \citet{Foley13}. 

\begin{figure}
\begin{center}
\leavevmode
\includegraphics[width=8cm]{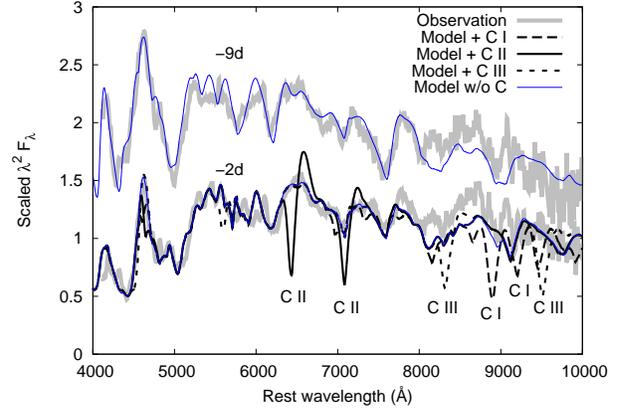}
\end{center}
\caption{Modified Model B spectra containing {\bf artificially enhanced} 
carbon features (thick solid and dashed lines) together with the $-2$d 
observed spectrum (grey curve) and the original, best-fit Model B spectrum without carbon (thin solid line). The observed and carbon-free model spectra 
at $-9$d are also shown for comparison.}
\label{fig:sp-carbon}
\end{figure}

As a first step, we simply added either C\,\begin{small}I\end{small}, C\,\begin{small}II\end{small} or C\,\begin{small}III\end{small} to the best-fit
Model B spectra to get carbon-enhanced models. The purpose of this step was
to check which carbon features should be visible under the physical 
conditions corresponding to Model B.     
Fig.~\ref{fig:sp-carbon} shows these carbon-enhanced models together with 
the carbon-free models and the $-2$d and $-9$d observed spectra. 
Note that the carbon features are artificially enhanced for making 
their identification easier. 

It is seen that large amount of C\,\begin{small}I\end{small} or C\,\begin{small}III\end{small} would cause observable features only at
wavelengths longer than $8000$ \AA\, where the observed spectra
have the lowest signal-to-noise and contamination from telluric lines
is the strongest. Thus, the reliable identification of these ions is
not possible from these spectra. 

However, the C\,\begin{small}II\end{small} 
$\lambda6580$ and $\lambda7234$ features offer better opportunity to
detect carbon in the optical spectra. There is a feature in both the 
$-9$d and the $-2$d observed spectra that could be C\,\begin{small}II\end{small} $\lambda7234$, as found by \citet{Foley13}. Note that this feature is 
originally modeled with high-velocity O\,\begin{small}II\end{small} 
in both Model A and B, which is not a secure identification (see below). 
Even though the simultaneous presence of the C\,\begin{small}II\end{small} 
$\lambda6580$ feature is not observed in the $-2$d spectrum, there may be
a weak feature at the expected position in the $-9$d spectrum. 

In order to test whether it is more appropriate to explain the 
feature around $7100$ \AA\ with C\,\begin{small}II\end{small} rather 
than O\,\begin{small}II\end{small}, we replaced O\,\begin{small}II\end{small}
with C\,\begin{small}II\end{small} in the best-fit $-9$d Model B spectrum,
and recomputed the fitting. The excitation temperature for carbon was set variable
to take into account the possibility for non-LTE excitations. It was found
that $T_{exc} = 20,000$ K (which is a factor of 2 higher than $T_{phot}$
at this phase) would indeed reduce the strength of the $\lambda6580$ feature relative
to that of the $\lambda7234$ one. 

\begin{figure}
\begin{center}
\leavevmode
\includegraphics[width=8cm]{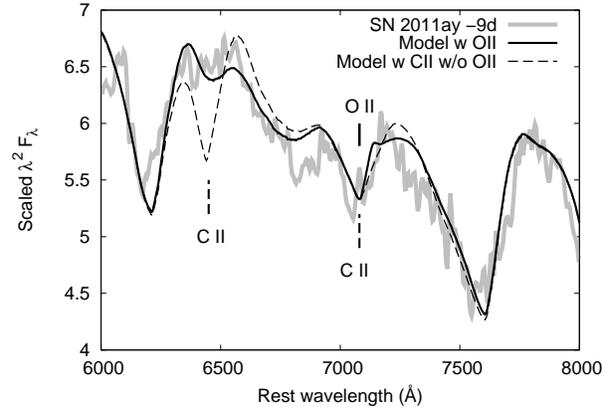}
\end{center}
\caption{Comparison of the modified Model B 
containing C II instead of high-velocity 
OII (dashed line) with the original Model B
(having OII, without CII,
thick solid line) and the observed $-9$d spectrum (grey curve). 
If CII $\lambda7234$ were present, $\lambda6580$
would appear much stronger than observed (see text for details).}
\label{fig:sp-carbon-new}
\end{figure}

In Fig.~\ref{fig:sp-carbon-new} the new 
carbon-enhanced model is compared with the original best-fit Model B spectrum
(without carbon) and the observations. It is seen that while the
C\,\begin{small}II\end{small} $\lambda7234$ feature is a similarly good fit 
to the observed spectrum as the high-velocity O\,\begin{small}II\end{small} 
feature in the original model, this is not true for the 
C\,\begin{small}II\end{small} $\lambda6580$ feature,
the latter being much stronger than the observed notch around $6400$ \AA.
Note that setting the excitation temperature of C\,\begin{small}II\end{small}
close to the photospheric temperature, $T_{phot} = 10,000$ K, would make
the $\lambda6580$ feature even stronger, enhancing the inconsistency 
between the carbon-enhanced model and the observations. 
Thus, if C\,\begin{small}II\end{small} were responsible for
the observed feature at $\sim 7100$ \AA\, then the $\lambda6580$ feature
should appear much stronger than observed.  

It is concluded that our SYNAPPS modeling does
not confirm the detection of C\,\begin{small}II\end{small} features in 
the pre-maximum optical spectra of SN~2011ay, and the same is true for
both C\,\begin{small}I\end{small} and C\,\begin{small}III\end{small}. 

The identification of the $\lambda 7100$ feature as high-velocity 
O\,\begin{small}II\end{small} in the early-phase spectra 
(see Fig.~\ref{fig:sp-modelions} and \ref{fig:minvel}) is also uncertain.
O\,\begin{small}II\end{small} is usually absent in the pre-maximum spectra
of SNe Ia, but it was found in other SNe Iax around maximum light, 
namely in 2008A and 2005hk, by \citet{McCully14a}. Given that our
models for SN~2011ay have photospheric temperatures not exceeding 
10,000 K (cf. Table~\ref{tab:synpar}), the appearance of 
O\,\begin{small}II\end{small} with  nearly the same strength as 
O\,\begin{small}I\end{small} is
not expected, at least in quasi-LTE conditions. \citet{Hatano99}
predicted equal optical depth for these two ions at $T \sim 12,000$
K, but note that they used $n_e \sim 5 \times 10^9$ cm$^{-1}$
for the electron density, a typical value for SNe Ia-s. If the
electron density in SN~2011ay were lower than this, then ionization
would be more effective at a given temperature, thus, the 
O\,\begin{small}II\end{small} features could become stronger
even at $T < 10,000$ K. Nevertheless, since it is based on only
a single feature, the presence of O\,\begin{small}II\end{small} 
in SN~2011ay cannot be proven unambiguously.

Regarding the intermediate-mass elements (IME),  Ca\,\begin{small}II\end{small} is close to the detection limit in the pre-maximum spectra, but its near-IR triplet (IR3) $\lambda 8579$ feature
becomes very strong by two weeks after maximum. Similar behavior 
is observed in overluminuous SNe Ia-91T \citep[see e.g.][]{Phillips92,Garavini04}, 
but those objects are characterized by very different light curve properties.

This kind of spectral evolution is the opposite of what is observed in normal SNe Ia 
\citep[see e.g.][]{howie_09ig}: in most SNe Ia the Ca\,\begin{small}II\end{small}
IR3 starts as a strong high-velocity feature (HVF) at $v \gtrsim 20,000$ km s$^{-1}$ 
at about two weeks before maximum, while low-Z elements are absent or very weak. By the time of
maximum light the Ca\,\begin{small}II\end{small} HVF disappears, 
while its photospheric component (PVF, $v \sim 10,000$ km s$^{-1}$) gets stronger. 
O\,\begin{small}I\end{small} $\lambda 7775$ can, sometimes, be detected as a
HVF at very early phases \citep{parrent_11fe}, but it is frequently
absent and its PVF shows up only after maximum light. 

It is seen in the spectra presented in this paper that SN~2011ay does not exhibit 
such kind of HVF for any ion at any phase observed. 
This also appears true for all other SNe Iax observed so far \citep{Foley13}.
It is interesting that the absence of Ca\,\begin{small}II\end{small} HVF is also 
characteristic for the low-velocity, 91bg subtype of SNe Ia 
\citep{Childress14, Silverman15}.

Unlike the Ca\,\begin{small}II\end{small} features, the Fe\,\begin{small}II\end{small} 
lines appear quite strong in SN~2011ay, even during pre-maximum.
The strong presence of Fe\,\begin{small}II\end{small} at such early phases 
were not observed in most ``normal'' SNe Ia spectra. 
Also, they were not identified in other pre-maximum Iax spectra, e.g. SN~2002cx \citep{Branch04} or SN~2005hk \citep{Sahu08}. 
Their strong appearance in SN~2011ay might be due to its cooler
photospheric temperature during pre-maximum: $T_{phot} \leq 10,000$ K (Table \ref{tab:synpar}). 
SNe Ia typically have $T_{phot} \gtrsim 10,000$ K before maximum. 
Concerning other SNe Iax, Fe\,\begin{small}II\end{small} features were found strong in the 
post-maximum spectra of SNe~2002cx \citep{Branch04} and 2005hk \citep{Sahu08},
when the photospheric temperature cooled below $\sim 10,000$ K.  
After maximum light, 
Fe\,\begin{small}II\end{small} and Ca\,\begin{small}II\end{small} features dominate the observed spectra of SN~2011ay (Fig.~\ref{fig:sp-modelall}), similar to SNe Ia.  

\subsection{Measuring velocities in SNe Iax spectra}\label{ana_vel}

One of the most intriguing properties of SNe Iax is their low expansion velocities 
\citep[e.g.][]{Branch04, Foley13}, which can, in extreme cases, be as low as $\sim 2000$ km s$^{-1}$
\citep{Stritzinger14}. This has been confirmed for multiple SNe Iax by spectral modeling with
SYNOW/SYNAPPS, i.e. similar methodology to that which has been applied in this paper. To estimate 
the expansion velocity for a bigger sample of SNe Iax, \citet{Foley13} used the projected 
Doppler-shift of the absorption minimum of the feature around $\lambda$6200 \AA, which was
interpreted as due to Si\,\begin{small}II\end{small} $\lambda$6355, similar to Type Ia SNe.
Based on this assumption, \citet{Foley13} concluded that the ``ejecta velocity'', $|v|$, 
defined this way is less than $8000$ km s$^{-1}$ for all SNe Iax. 
All SNe Iax spectra studied to date indeed show signs for significantly lower expansion
velocities than most SNe Ia which usually have $v_{\rmn{SiII}} > 10,000$ km s$^{-1}$ 
around maximum light \citep[e.g.][]{Blondin06}, however, we show below that such ``quick-look''
velocity estimates based on single spectral features may lead to significantly under- or
overestimated velocities for SNe Iax, which have much more complex spectra than SNe Ia.



\begin{figure}
\begin{center}
\leavevmode
\includegraphics[width=8cm]{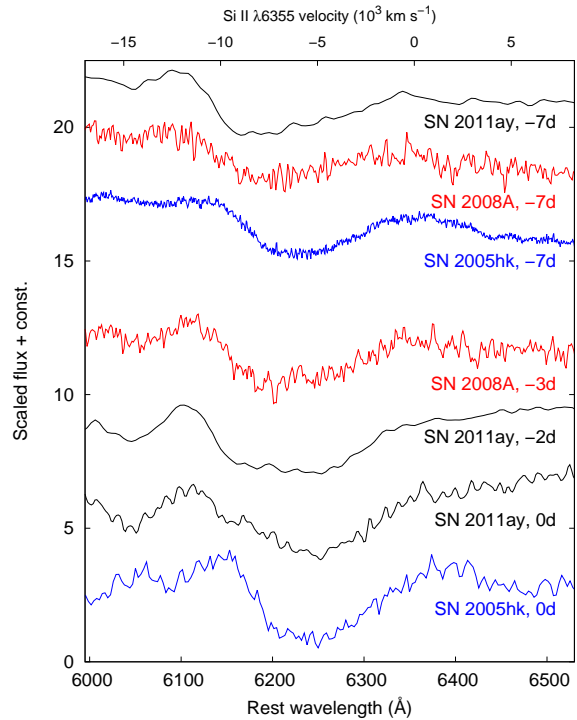}
\end{center}
\caption{Comparison of the near-maximum spectra of three different SNe Iax in the region of 6000--6500 \AA: SN~2005hk \citep{Phillips07}, 
SN~2008A \citep{Blondin12}, and SN~2011ay \citep[present work and][]{Foley13}. All epochs are given relative to V maxima. 
The spectra of SNe 2005hk and 2008A were downloaded from WISeREP \citep{Yaron12}. A velocity scale relative to Si II $\lambda$6355 is also shown on the upper horizontal axis.}
\label{fig:6355_sne}
\end{figure}

We focus on the feature appearing around $\sim \lambda$6200 \AA\ in the pre-maximum spectra of
SN~2011ay. Fig.~\ref{fig:6355_sne} shows this feature before and around maximum light
compared to two other SNe Iax observed at similar phases. It is seen that this feature is 
broad and shallow. In this respect, SN~2011ay and some other SNe Iax look similar to 
SNe Ia-91T \citep{Branch04, Foley13}, i.e. they usually show weaker Si\,\begin{small}II\end{small} $\lambda$6355 lines compared to normal Ia-s. However, the post-maximum spectral evolution of the
two classes of SNe are markedly different. 

The first issue with the velocity measurement from this feature is related to its line strength.
\citet{JB90} studied the effects of line strength, ejecta density and other parameters on the Doppler-shift of the absorption minimum of P Cygni features in SN atmospheres. They showed that for features
having optical depth $\tau \sim 1$ at the photosphere, the absorption minimum forms very
close to $v_{phot}$. That is why the velocity from the Si\,\begin{small}II\end{small} 
$\lambda$6355 feature
gives so reasonable photospheric velocities in SNe Ia: this feature is strong and relatively
unblended in SNe Ia \citep[e.g.][]{parrent_11fe}. 
For weak features having $\tau < 1$, however, resonant scattering tends
to dominate over absorption, which may shift the absorption minimum significantly below 
$v_{phot}$. On the contrary, much stronger features, like Ca\,\begin{small}II\end{small}, 
tend to show absorption minima at velocities much higher than $v_{phot}$, as expected.

Fig.~\ref{fig:6355_sne} suggests that SN~2011ay had similar photospheric velocity to 
SN~2008A, and both of them expanded faster than SN~2005hk. Using SYNAPPS, 
\citet{McCully14a} derived $\sim 7000$ and $\sim 8500$ km s$^{-1}$ for 2005hk and 2008A, respectively, while \citet{Foley13} estimated $\sim4500$ and $\sim6400$ km s$^{-1}$ from the $\lambda$6200 feature.
The $\sim 2000$ km s$^{-1}$ difference between the single-feature and the spectrum-modeling velocity illustrates the issue of the velocity measurement in SNe Iax spectra. For SN~2011ay we find
a similar conflict: at maximum light the single-feature velocity estimate gives $\sim 5600$ km s$^{-1}$
\citep{Silverman11,Foley13}, while our Model A predicts $v_{phot} \sim 9000$ km s$^{-1}$ (with an uncertainty of at least $\sim 1000$ km s$^{-1}$). It seems that SYNAPPS modeling tend to produce higher 
velocities than the single-feature ``quick-look'' estimates.


\begin{figure}
\begin{center}
\leavevmode
\includegraphics[width=8cm]{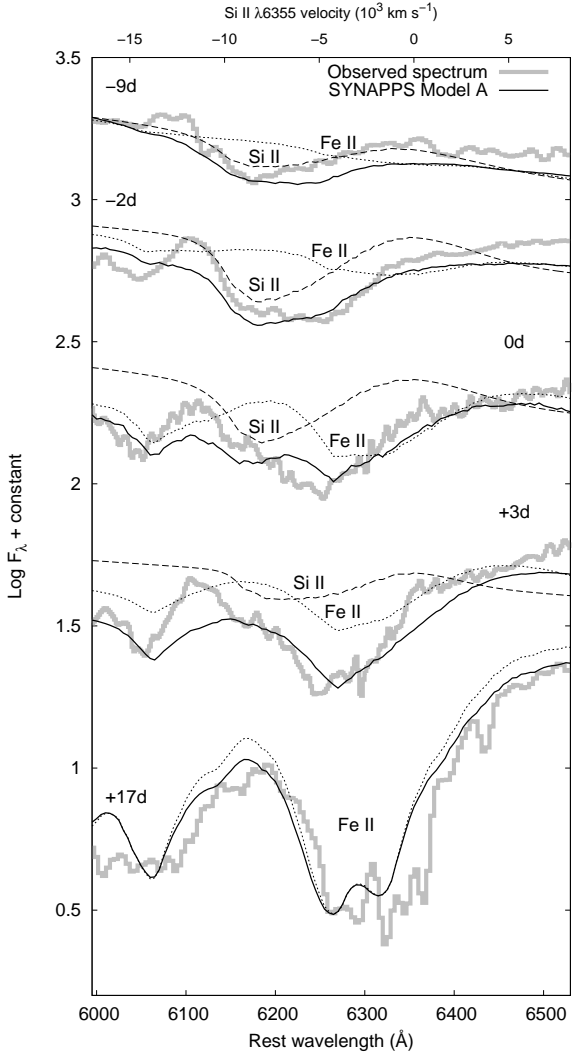}
\end{center}
\caption{Spectral evolution of SN~2011ay between 6000--6500 \AA. The observed spectrum (thick
gray line) is plotted together with the SYNAPPS model from Model A (solid line) as well
as single-ion contributions from Si II (dashed line) and Fe II (dotted line). 
Fe II $\lambda$6456 starts blending with Si II $\lambda$6355 a few days before maximum, 
and it becomes the dominant feature post-maximum. A velocity scale relative to 
Si II $\lambda$6355 is also shown on the upper horizontal axis.}
\label{fig:6355-A}
\end{figure}

\begin{figure}
\begin{center}
\leavevmode
\includegraphics[width=8cm]{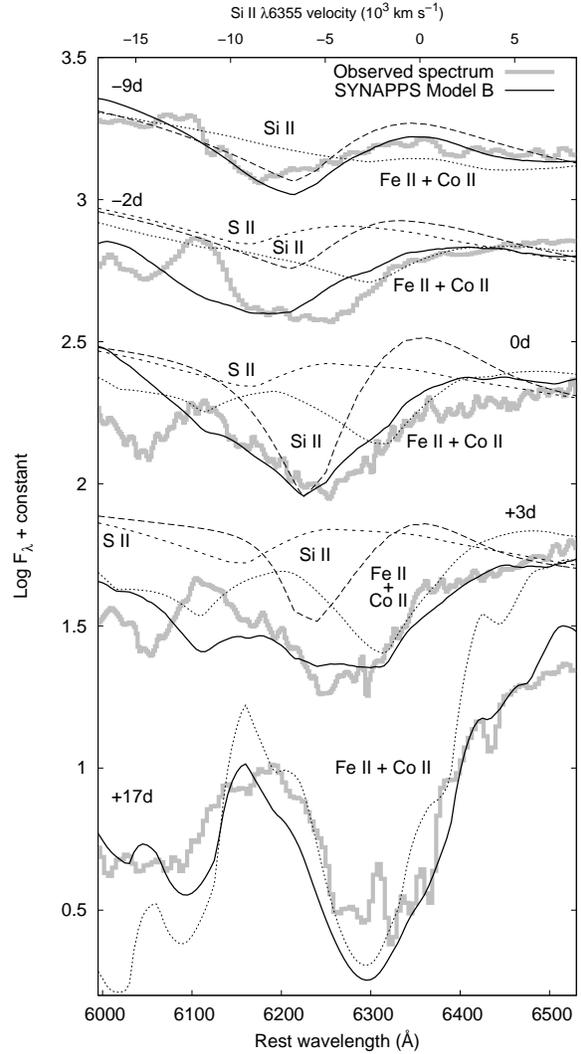}
\end{center}
\caption{The same as Fig.~\ref{fig:6355-A} but showing Model B, and adding the single-ion
contributions from S II and Co II. Complex blending between these features results
in nearly the same observed spectra, even though the photospheric velocities are $\sim 3000$
km s$^{-1}$ less than in Model A.}
\label{fig:6355-B}
\end{figure}

However, there is an even bigger issue with the velocity measurement in SNe Iax spectra. 
This is the severe blending across the whole optical regime, which prevents any
unique, secure line identification, as was detailed in the previous section. 
This affects the $\lambda$6200 feature, further plaguing the velocity determination.  

Fig. \ref{fig:6355-A} again zooms in the 6000 - 6500 \AA\ regime while showing the observed spectral 
evolution of SN~2011ay together with the model spectra by Model A (i.e. the one having
higher $v_{phot}$). From the single-ion contributions it is seen that the broad $\lambda$6200 feature
is due to blending between Si\,\begin{small}II\end{small} $\lambda$6355 and Fe\, \begin{small}II\end{small} $\lambda$6456, even during the pre-maximum phases.  
Similar blending is also visible in other SNe Iax, as found in  
2002cx at +7d \citep[$T_{BB} \sim 9000$ K,][]{Li03,Branch04}, 
2005hk at -1d \citep[$T_{BB} \sim 9000$ K,][]{Phillips07,Sahu08}, 
2008A at -3d \citep{McCully14a}, 
and 2010ae at +16d \citep{Stritzinger14}.
This blending broadens the observed profile and shifts the middle of the feature toward lower velocities, which may likely explain the discrepancies between the ``quick-look'' and the
modeled velocities mentioned above. 

Moreover, this is not yet the full story, because, as illustrated in Fig.~\ref{fig:6355-B},
blending between the different single-ion features in this regime may result in nearly the same
observed spectrum, even when the photospheric velocity is close to $\sim 6000$ km s$^{-1}$, i.e.
close to what the ``quick-look'' velocity estimate predicts. Thus, even when using a sophisticated
parametrized spectrum modeling code and putting in all possible features that likely contribute to
the observed spectrum, the resulting model may still be ambiguous: if the observed spectrum consists
of broad features and almost all of them are complex blends, then the photospheric velocity becomes rather ill-constrained. 

A good example for this ambiguity is SN~2011ay: as shown in the previous 
section, its observed spectra
can be explained either with $v_{phot} \sim 6000$ or $\sim 9000$ km s$^{-1}$ almost equally well.
This and the appearance of the detached features (those having higher velocities than $v_{phot}$)
in the lower velocity Model B might also be a hint for a multiple velocity structure in the ejecta,
i.e. a disk-like or a jet-like configuration. Investigations of such models are beyond the scope
of this paper, and might be possible only by having more extensive data coverage, both in wavelength
(i.e. by adding infrared and/or UV spectra) and in time.

\subsection{Spectral energy distributions and light-curve modeling}\label{ana_lcmod}

\begin{figure*}
\begin{center}
\leavevmode
\includegraphics[width=8cm]{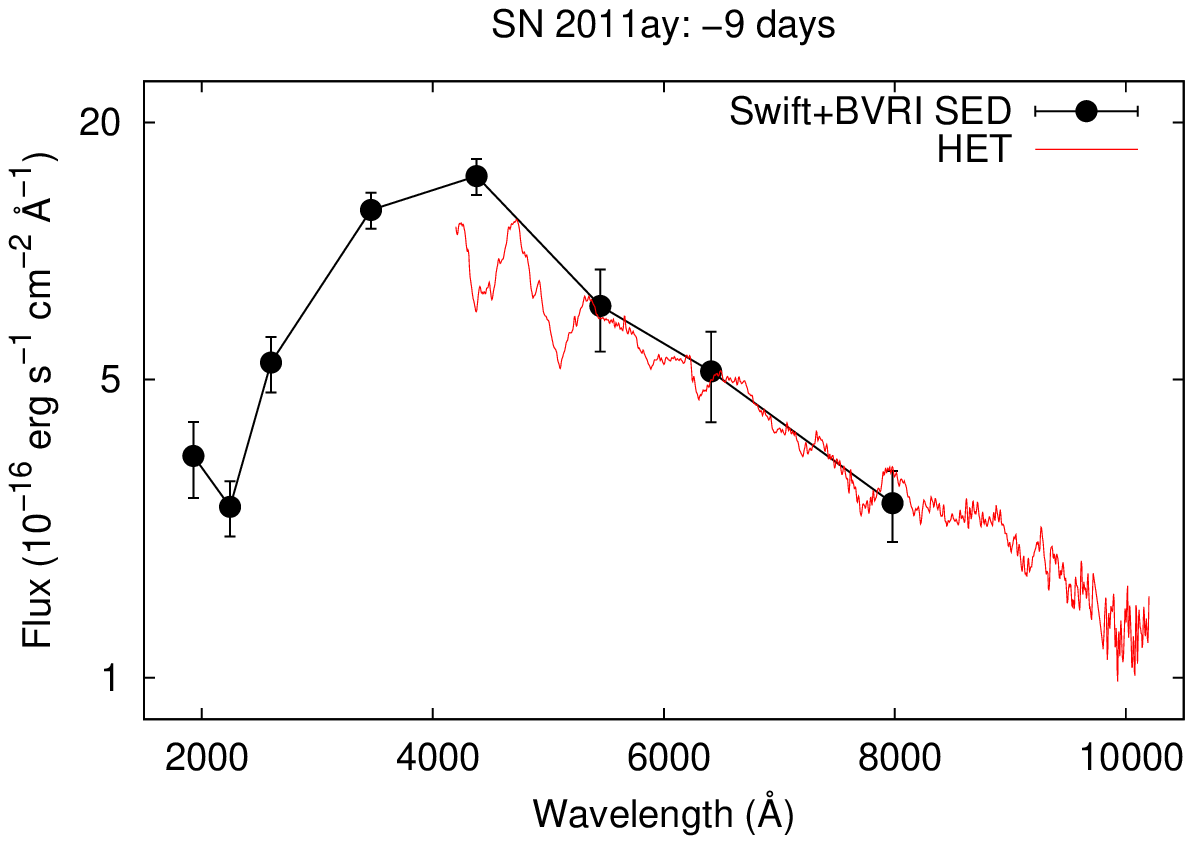}
\includegraphics[width=8cm]{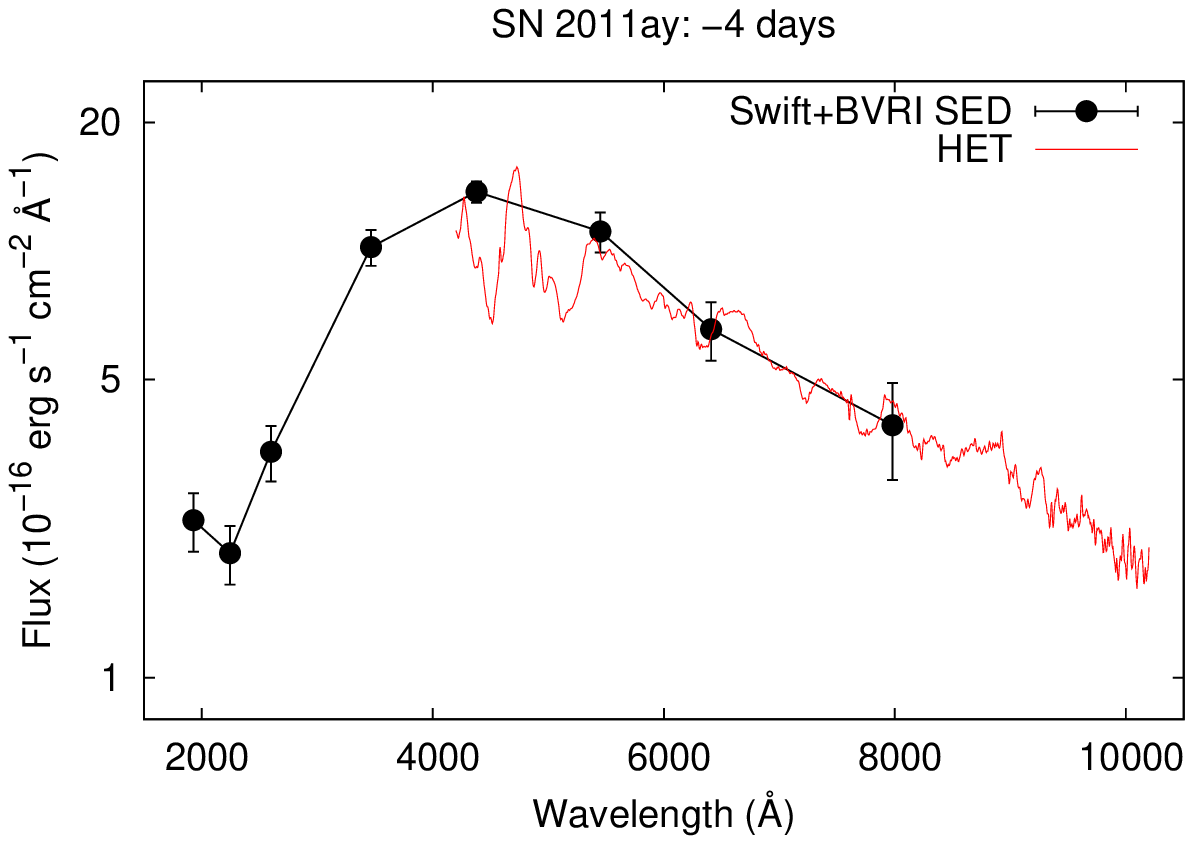}
\includegraphics[width=8cm]{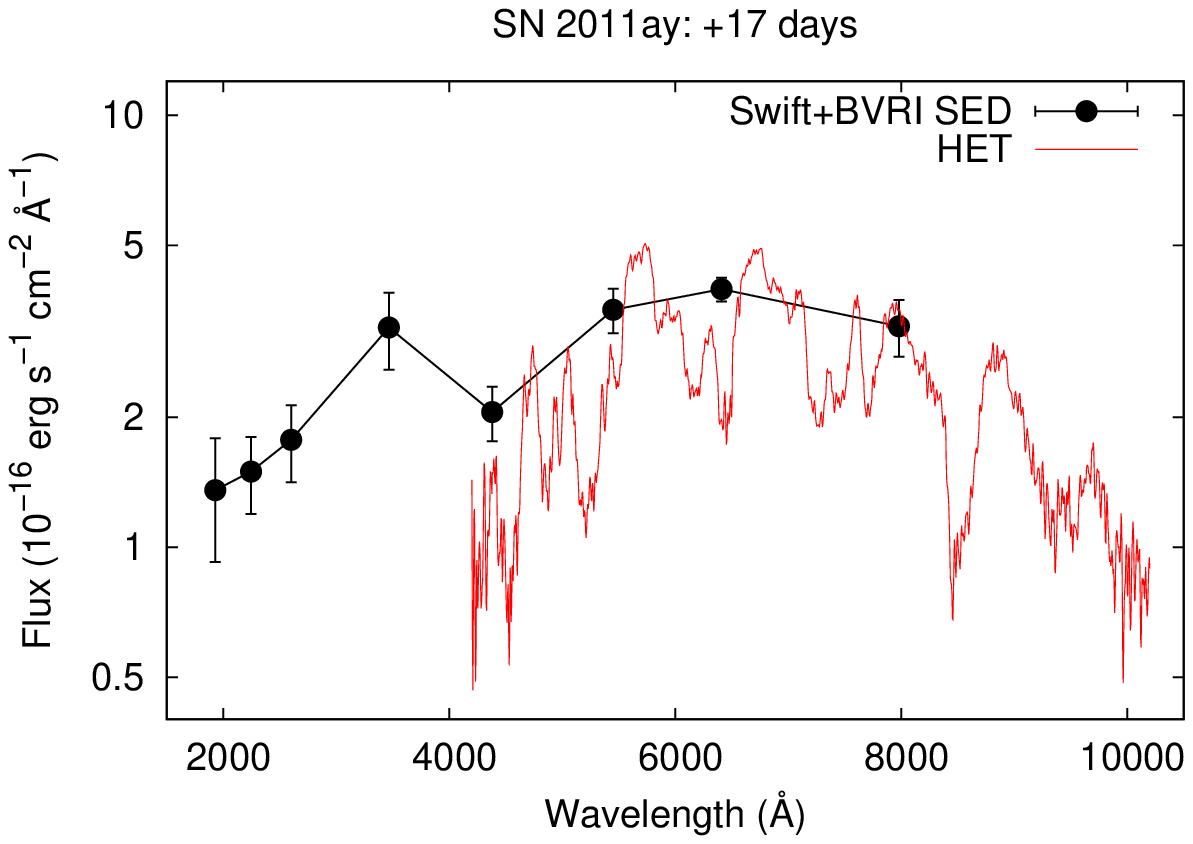}
\end{center}
\caption{Scaled HET spectra and the UV-optical SEDs of SN~2011ay at different epochs relative to V maximum: -9 days (2011 March 24, top left), -4 days (2011 March 29, bottom left), +17 days (2011 April 19, bottom right).}
\label{fig:sed_sp}
\end{figure*}

To calculate the SEDs of SN~2011ay, we converted the BVRI and {\it Swift/UVOT} magnitudes to $F_{\lambda}$ fluxes using the calibration of 
\citet{Bessell98}. The fluxes were dereddened using the galactic reddening law parametrized by \citet{Fitzpatrick07} 
assuming $R_V = 3.1$ and adopting $E(B-V)=0.081$ mag (\S 3.1.). The combined UV-optical SEDs are compared with the three 
HET spectra in Figure \ref{fig:sed_sp}. After correcting for the small uncertainties in the absolute flux 
calibration of HET data, the spectra and the broadband SEDs match very well. 
Trustworthiness of {\it Swift} data at +17 days are lower, because SN~2011ay was already quite faint in the UV
at that time.

The quasi-bolometric light curve was derived by integrating the dereddened $F_{\lambda}$ values of the combined
UV-optical SEDs against wavelength. The long-wavelength contribution (not covered by these data) was estimated 
by fitting a Rayleigh-Jeans tail to the red end of the observed SEDs, and integrating it to infinity. 
In addition, the light curve was supplemented by two published KAIT-measurements: a non-detection 
on March 1 (2\,455\,621.7 JD), and the unfiltered magnitude at the discovery \citep{Blanchard11}. 
The luminosity calculated from the latter was used as a lower limit at that epoch. 
Finally, the integrated fluxes were converted to luminosities using $D=86.9$ Mpc (see \S 1).

We applied a generalized analytic light curve model published by \citet{Chatzopoulos09,Chatzopoulos12} to derive 
the main parameters of the SN. This model is based on the radioactive decay diffusion model of \citet{Arnett82} 
and its generalized form by \citet{Valenti08}, but also takes into account the gamma-ray leakage from the ejecta. 
In this model the output luminosity can be expressed as 

\begin{eqnarray}
L(t) = M_{\rmn{Ni}} \mathrm{e}^{-x^2} \left[ \left( \epsilon_{\rmn{Ni}} - \epsilon_{\rmn{Co}} \right) \int_0^x 2 z \mathrm{e}^{z^2-2zy}\,\mathrm{d}z + \epsilon_{\rmn{Co}} \right. \nonumber \\ 
\left. \times \int_0^x 2 z \mathrm{e}^{z^2-2yz+2zs}\,\mathrm{d}z \right] \left( 1 - \mathrm{e}^{-At^{-2}} \right) ,
\label{eq:cwv}
\end{eqnarray}

\noindent where $x = t/t_{\rmn{m}}$, $t_{\rmn{m}}$ is the effective diffusion time (roughly equal to the rise time to maximum), 
$y = t_{\rmn{m}}/(2t_{\rmn{Ni}})$ with $t_{\rmn{Ni}} = 8.8$ days, 
$s = t_{\rmn{m}}(t_{\rmn{Co}}-t_{\rmn{Ni}})/(2t_{\rmn{Co}}t_{\rmn{Ni}})$ with $t_{\rmn{Co}} = 111.3$ days, 
$M_{\rmn{Ni}}$ is the initial nickel mass, 
$\epsilon_{\rmn{Ni}} = 3.9 \times 10^{10}$ erg\,s$^{-1}$\,g$^{-1}$ and 
$\epsilon_{\rmn{Co}} = 6.8 \times 10^9$ erg\,s$^{-1}$\,g$^{-1}$ 
are the energy generation rates due to Ni- and Co-decay. 

The last term of Eq. (\ref{eq:cwv}) describes the amount of gamma-ray leaking. 
Assuming a spherical uniform density ejecta with radius $R = vt$, homologous expansion 
(due to $t^{-2}$ scaling), and the Ni/Co confined  in the center, yields 
$A = \left(3 \kappa_{\gamma} M_{\rmn{ej}} \right)/ \left(4 \pi v^2 \right)$, 
where $\kappa_{\gamma}$ is the gamma-ray opacity, $M_{\rmn{ej}}$ is the ejecta mass.
Note that we used the dimensionless form of 
$\left( 1 - \exp \left(-A_{\rmn{norm}} \cdot t_{10}^{-2}\right) \right)$, 
where $t_{10} \equiv$ $t$/10 days.

\begin{figure}
\begin{center}
\leavevmode
\includegraphics[width=8cm]{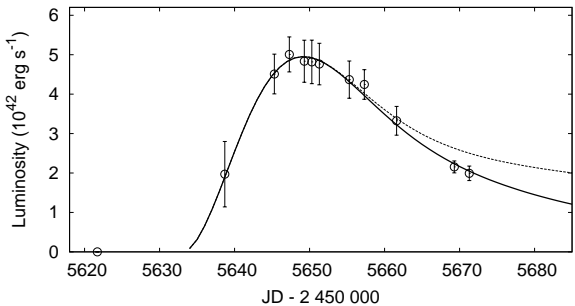}
\end{center}
\caption{Quasi-bolometric light curve of SN~2011ay with the best-fitting Arnett-model (solid line) and the one with assuming full gamma-ray trapping (dashed line). 
The first two luminosities were calculated from the measurements of \citet{Blanchard11}, see the text for details.}
\label{fig:lc_model}
\end{figure}

The quasi-bolometric light curve and the best-fitting model are shown in 
Figure \ref{fig:lc_model}. The corresponding parameters are presented in Table \ref{tab:lcpar}.
Both the nickel mass ($\sim 0.22$ $M_\odot$) and the rise time ($\sim 14$ d) are lower
than the usual values for SNe Ia ($\sim 0.6$ $M_\odot$ and $\sim 18$ d, respectively) , 
similar to the conclusion by \citet{Foley13}. 
The short rise time implies low ejecta mass. 
%
%
%
Following the method applied by \citet{Foley09} for SN~2008ha and assuming that the ejecta of SN~2011ay has the 
same mean opacity as that of a normal SN Ia, we have

\begin{equation}
E_{\rmn{kin}} / E_{\rmn{kin,Ia}} = \left(\frac{v}{v_{\rmn{Ia}}}\right)^3 \left(\frac{t_{\rmn{rise}}}{t_{\rmn{rise,Ia}}}\right)^2 
\end{equation}

\noindent and

\begin{equation}
M_{\rmn{ej}} / M_{\rmn{ej,Ia}} = \frac{v}{v_{\rmn{Ia}}} \left(\frac{t_{\rmn{rise}}}{t_{\rmn{rise,Ia}}}\right)^2 .
\end{equation}

\noindent Adopting $v \sim 9,300$ km s$^{-1}$, as suggested by
the higher-velocity Model A (\S 3.2) and $t_{\rmn{rise}}$ = 14 days together with the reference values of 
$E_{\rmn{kin,Ia}}$ = 10$^{51}$ erg, 
$M_{\rmn{ej,Ia}}$ = 1.4 $M_{\odot}$, 
$v_{\rmn{Ia}}$ = 10\,000 km s$^{-1}$ \citep[][]{Foley09} 
and the average value of $t_{\rmn{rise,Ia}}$ = 18.0 days \citep[see][]{Ganeshalingam11}, 
these formulae result in 
$E_{\rmn{kin}}$ = 0.5$\cdot$10$^{51}$ erg and 
$M_{\rmn{ej}}$ = 0.8 $M_{\odot}$ 
as the kinetic energy and the ejecta mass of SN~2011ay, respectively. 
Alternatively, using $v \sim 6000$ km s$^{-1}$ from the lower-velocity Model B results in $M_{\rmn{ej}} \sim 0.5 M_{\odot}$ and $E_{\rmn{kin}} = 0.1 \cdot$10$^{51}$ erg.
The ejecta mass and expansion energy estimated this way 
are consistent with previous findings for SNe Iax. The lower nickel mass, $M_{\rmn{Ni}} \sim 0.2$ $M_\odot$ as calculated from the Arnett-model, also suggests that the explosion, if indeed thermonuclear, may be less energetic that in Type Ia-s.  

\begin{table}
\begin{center}
\caption{Physical parameters determined from the modeling of the bolometric light curve.}
\label{tab:lcpar}
\begin{tabular}{lll}
\hline
\hline
$t_0$ (JD) & 2\,455\,633.0 $\pm$ 1.5 & Date of explosion\\
$t_{\rmn{rise}}$ (d) & 14 $\pm$ 1 & Rise time to maximum \\
$M_{\rmn{Ni}}$ ($M_{\odot}$) & 0.225 $\pm$ 0.010 & Initial $^{56}$Ni mass\\
$A_{\rmn{norm}}^a$ & 25 $\pm$ 7 & $\gamma$-ray leakage parameter\\
\hline
\end{tabular}
\end{center}
\smallskip
{\bf Notes.} $^a$The parameter connected to gamma-ray leaking (see Eq. \ref{eq:cwv} and text for details). 
\end{table}

\section{Discussion and conclusions}\label{conc}

Based on photometric and spectroscopic data obtained during the early phases 
we carried out a detailed analysis on SN 2011ay, one of the members of the recently defined class of SNe Iax. 
The spectra as well as the light and color curves are similar to those of other objects belonging to this group 
\citep[see][and references therein]{Foley13}. 

We calculated model spectra with the parametrized modeling code SYNAPPS to get more information on the spectral evolution and the physical properties of the ejecta.
As presented in Section \ref{ana_sp}, the spectral characteristics of SN~2011ay are
basically similar to those of other SNe Iax. 
We found that Fe\,\begin{small}II\end{small} features appear even before 
maximum, which can be explained with the relatively lower photospheric temperature ($T_{phot} \sim$ 8,000 K) of the ejecta. 

The presence of Fe\,\begin{small}II\end{small} and other features 
that cause severe blending across the entire visible spectral domain makes 
the direct analysis of such spectra very difficult and ambiguous. We found
that it is possible to fit most of the broad features of SN~2011ay by two
models (Model A and B) that have similar chemical composition but very
different photospheric velocity ($\sim 9,300$ km s$^{-1}$ and 
$\sim 6,000$ km s$^{-1}$ at maximum, respectively).  

The effect of strong blending also makes the ``quick-look'' velocity
estimates uncertain.  
We showed in Section \ref{ana_vel} that measuring the absorption minimum of the 
$\lambda$6200 feature and assuming that it is due to purely 
Si\,\begin{small}II\end{small} $\lambda$6355 might underestimate $v_{phot}$, because
of the effect of blending with Fe\,\begin{small}II\end{small} and 
Co\,\begin{small}II\end{small}. This does not question the fact that
SNe Iax generally have lower expansion velocities than SNe Ia, but
the non-uniqueness of the spectrum modeling fits presented in this paper
makes the velocities of such events rather ill-constrained. 


Fitting a radiative-diffusion model (taking into account gamma-ray leaking) to the quasi-bolometric
light curve of SN~2011ay resulted in physical parameters for the rise time to maximum, ejecta mass, 
initial nickel mass and kinetic energy that are all below their mean value for SNe Ia, as pointed
out by \citet{Foley13}. Although SN~2011ay seems to be close to the upper limits of these parameters,
it clearly belongs to the recently defined class: the values of $^{56}$Ni mass and total ejecta mass 
(0.225$\pm$0.010 $M_{\odot}$ and $\sim$0.8 $M_{\odot}$, respectively) are similar to the parameters 
of other, bright (M$_V >$ -18 mag) SNe Iax \citep[SNe 2005hk, 2008A, 2009ku; see][]{Foley13,McCully14a}.
Note that in addition to the criteria defined by \citet{Foley13},
the faster decline rate of the bolometric light curve, indicated by the relatively low 
value of the gamma-ray trapping
parameter $A_{norm}$ (see Table~\ref{tab:lcpar}), may also be characteristic for SNe Iax.   

The results of the light curve modeling strengthen the picture that the progenitor of SN~2011ay 
is likely an incompletely exploded white dwarf, which seems to be the best explanation 
of this kind of SNe to date. The presence of Si\,\begin{small}II\end{small} and 
S\,\begin{small}II\end{small} features in the early spectra of SN~2011ay in Figure \ref{fig:sp-modelall} may be signs of thermonuclear explosion, which might give additional support for the
white dwarf explosion scenario.

These results, as well as other detailed studies of single events, may help us to understand better the properties of type Iax SNe. 
However, there are still a lot of open questions about the true nature and origin of this class of 
stellar explosions.

\section*{Acknowledgments} 

This work has been supported by the Hungarian OTKA Grants NN107637, K104607, and K83790. 
TS is supported by the OTKA Postdoctoral Fellowship PD112325.
JCW's Supernova group at the UT Austin is supported by 
NSF Grant AST 11-09881 grant. KS and JB are supported by the Lend\"ulet-2009 Young Researchers’ Program of the Hungarian Academy of Sciences. 
JMS is supported by an NSF Astronomy and Astrophysics Postdoctoral Fellowship under award AST-1302771.
KT is supported by the Gemini-CONICYT Fund, allocated to the project N$^o$ 32110024.
An anonymous referee provided a thorough report that helped us to
improve the previous version of this paper. His/her work is truly appreciated.
We also thank R.J. Foley for providing the Lick/Kast spectra of SN 2011ay.

The Hobby-Eberly Telescope (HET) is a joint project of the University of Texas at Austin, the Pennsylvania State University, 
Stanford University, Ludwig-Maximilians-Universit\"at M\"unchen, and Georg-August-Universit\"at G\"ottingen. 
The HET is named in honor of its principal benefactors, William P. Hobby and Robert E. Eberly.
The Marcario Low Resolution Spectrograph is named for Mike Marcario of High Lonesome Optics 
who fabricated several optics for the instrument but died before its completion. 
The LRS is a joint project of the Hobby-Eberly Telescope partnership and the Instituto de Astronom\'ia 
de la Universidad Nacional Aut\'onoma de M\'exico.
We acknowledge the thorough work of the HET resident astronomers, Matthew Shetrone, Stephen Odewahn, John Caldwell
and Sergey Rostopchin during the acquisition of the spectra.  

This research has made use of the NASA/IPAC Extragalactic Database (NED) which is operated by the Jet Propulsion Laboratory, California Institute of Technology, under contract 
with the National Aeronautics and Space Administration. We acknowledge the availability of NASA ADS services.

\label{lastpage}

\end{document}